\documentclass[12pt]{article}
\usepackage{arxiv}
\usepackage[utf8]{inputenc} 
\usepackage[T1]{fontenc}    

\usepackage{url}            
\usepackage{booktabs}       
\usepackage{amsfonts}       
\usepackage{nicefrac}       
\usepackage{microtype}      
\usepackage{graphicx}
\usepackage{amsmath}
\usepackage[english]{babel}
\usepackage{csquotes}
\usepackage{hyperref}
\usepackage{fancyhdr,lipsum}
\graphicspath{{Images/}}
\usepackage{tabularx}
\title{ Comparison of Deep learning models on time series forecasting : a case study of Dissolved Oxygen Prediction}
\usepackage{setspace}
\author{
  Hongqian Qin \\
  University of Pennsylvania\\
  Philadelphia, PA, 19104 \\
  \texttt{hongqian@upenn.edu} \\
}

\begin{document}
\setlength{\headheight}{16pt}
\maketitle
  
\begin{abstract}
Deep learning has achieved impressive prediction performance in the field of sequence learning recently. Dissolved oxygen prediction, as a kind of time-series forecasting, is suitable for this technique. Although many researchers have developed hybrid models or variant models based on deep learning techniques, there is no comprehensive and sound comparison among the deep learning models in this field currently. Plus, most previous studies focused on one-step forecasting by using a small data set. As the convenient access to high-frequency data, this paper compares multi-step deep learning forecasting by using walk-forward validation. Specifically, we test Convolutional Neural Network (CNN), Temporal Convolutional Network (TCN), Long Short-Term Memory (LSTM), Gated Recurrent Unit (GRU), Bidirectional Recurrent Neural Network (BiRNN) based on the real-time data recorded automatically at a fixed observation point in the Yangtze River from 2012 to 2016. By comparing the average accumulated statistical metrics of root mean square error (RMSE), mean absolute error (MAE), and coefficient of determination in each time step, We find for multi-step time series forecasting, the average performance of each time step does not decrease linearly. GRU outperforms other models with significant advantages.
 
\end{abstract}


\section{Introduction}
 The quick development of industrialization and urbanization generates threats of water security. Long-term water contamination,  sudden pollution incidents,  plus the complicated distribution of sewage outlets and water intakes, put high pressure on both the aquaculture and the drinking water supply. To prevent such issues, we need a reliable model to predict the trend of change in water quality parameters. Dissolved oxygen (DO), a necessity for all aquatic animals, is considered an essential indicator of the water quality parameters. This paper will specifically discuss the univariate time series forecasting.\\
 Because of the development of techniques on real-time monitoring plus the increased investment in water quality management from the government, the approach of getting the data of water quality parameters has changed from low-frequency manual collection with low precision to high-frequency automatic collection with high accuracy. The relative cheap access to a tremendous amount of data shifts the focus on DO forecasting from limited data to big non-obvious seasonal data. The tech being applied also develops from stochastic models to non-linear models, and then to deep learning models. The history of the development of time series prediction is as follows:
 \begin{enumerate}
  \item Stochastic models, such as Autoregressive Integrated Moving Average (ARIMA), Kalman filter, and their variants, were popular on time series forecasting. They guarantee to converge to a unique optimal solution \textbf{[1]}, and they are simple for understanding and implementation. However, it cannot handle the non-linear relationship among the data.
  \item As a data-driven and self-adaptive non-linear model \textbf{[2][3]},  Artificial Neural Networks (ANNs) has the advantages that no need of statistical hypothesis for model application or a specific model form \textbf{[4]}. It can approximate any continuous function with impressive performance \textbf{[5]}. Although ANNs and its variants have become the hotspot on time series prediction in recent decades, it suffers from the challenges of the selection of proper parameters, and its final result cannot guarantee to be converged as the globally optimal.
  \item Another technology, Support Vector Regression (SVR), receives growing attention in the field of time series forecasting. It is the application of Support Vector Machine (SVM) on regression analysis. It covers advantages both from stochastic methods and ANNs. It is good at handling non-linear and non-stationary relationships among data without the request of statistical hypothesis \textbf{[1]}, and its solution is always globally optimal \textbf{[4]}. SVR has been proven to outperform neural networks \textbf{[1]}. But its main drawback is when dealing with the large training set, it needs enormous computation, which makes it computationally expensive \textbf{[6]}.
   \item Currently, deep learning has achieved the state-of-the-art in many time series forecasting tasks. As a neural network architecture with more hidden layers and more neurons, deep learning can automatically extract and handle more complex features and dependency relationships. For example, LSTM can catch both long-term and short-term dependencies in a sequence; BiRNNs can build the model automatically from both past information and future information. Besides that, deep learning is also able to deal with multiple inputs and outputs \textbf{[7]}. While its drawback is to tune the hyperparameters. 
\end{enumerate}

The remaining of this paper is structured as follows: Section 2 introduces the related studies on dissolved oxygen or water quality parameters forecasting. Section 3 describes the five deep learning models we test in this study. We show the detail of the experiment in Section 4. Section 5 includes our discussion of limitations and future work. Finally, we conclude in Section 6.

\section{Related Work}

In the field of DO or water quality parameter prediction, some researchers were interested in ANNs, their variants, or hybrid models. Sirilak Areerachakul \textbf{[8]}, Najah \textbf{[9]}, Majer \textbf{[10]}, Diamantopoulou \textbf{[11]}, Emamgholizadeh\textbf{[12]} experimented with ANNs; Yan Jiang \textbf{[13]} employed BP neural network; Chen Siyu \textbf{[14]} amalgamated an improved artificial bee colony (IABC) algorithm and BP neural network; Zhenbo Li \textbf{[15]} blended gray model with BP neural network; Yalin Liu \textbf{[16]} tried fuzzy neural network; Ahmed \textbf{[17]} combined a feed-forward neural network (FFNN) model and a radial basis function neural network (RBFNN) model; L.Zhang \textbf{[18]}, W.Deng \textbf{[19]} integrated ARIMA and RBFNN. 

 Some other researchers studied water quality parameter forecasting using SVR, SVM, their variants, or hybrid models. Ji Xiaoliang \textbf{[20]} found SVM worked well in hypoxic river systems; M.Liu \textbf{[21]} utilized improved SVM in nonpoint source polluted river; Shuangyin Liu \textbf{[22][23]} mixed Wavelet Analysis (WA) and Least Squares Support Vector Regression (LSSVR). Y.Xiang \textbf{[24]} utilized the least squares support vector machine (LS-SVM) combined with particle swarm optimization (PSO); Chakraborty \textbf{[25]} built a hybrid model integrating regression tree and SVM. Chen Li \textbf{[26]} proposed the model merging the least squares support vector machine (LSSVR), BP neural network, and grey model (GM).

 Although all authors claimed their methods achieved great results, what the models they compared were the original ARIMA or ANNs, and the specific settings made it hard to generalize to other water systems. Recently, as the big data era, the cost of access to an enormous amount of data has decreased, while the computational power has increased. That provides the opportunity for deep learning models to outperform the conventional time series models. CNN \textbf{[27]}, LSTM \textbf{[28-31]} were tried recently and were proved get better prediction capability than traditional machine learning or statistical methods in the field of time series forecasting problems. Another approach called gated recurrent unit (GRU)  was created as a modification of LSTM with a simpler structure \textbf{[32]}. It was proved to have better prediction performance than  RNN \textbf{[33]} and was applied widely on sequence prediction problems \textbf{[34][35]}.  Schuster and Paliwal \textbf{[36]} proposed bidirectional RNN to enable the information passed both in the forward direction and backward direction. Merging the bidirectional RNN with LSTM and GRU was originally designed for problems like text and speech recognition. Recently however, researches showed BiRNN also worked well on time series data prediction problems \textbf{[37-39]}. 
 
Plus, the machine learning department of Carnegie Mellon University introduced a generic convolutional architecture, temporal convolutional network \textbf{[40]}, for sequence modeling. That was developed based on CNN by applying dilated convolution and residual block. They claimed it outperformed LSTM in a broad sequence modeling tasks. 

To the best of our knowledge, the GRU, BiRNN, and TCN have not been researched on the prediction of DO. In this paper, we compare and analyze the prediction performance of CNN, LSTM, GRU, BiLSTM, BiGRU, and TCN. The following sections of this paper are as follows: Section 2 introduces the theories and architectures of these methodologies. Section 3 describes the experiment and explains its result. Section 4 provides the conclusion of our analysis. 

\section{Methodologies}
\label{sec:headings}

Because of the fast development of computational power while the slower improvement of RAM, statistical models have the power to optimize parameters fast while they have to be memory efficient simultaneously. Hence, the focus of statistics and machine learning are converting to methods with more nonlinearity like deep networks from traditional methods like linear models and kernel methods \textbf{[41]}. In this section, we will introduce five types of deep learning methods on time series prediction. They are LSTM, GRU, BiRNN, CNN, and TCN. 

\subsection{Long Short Term Memory (LSTM)} 
LSTM has become the state-of-the-art model for many sequence problems such as analysis of machine translation \textbf{[42]}, audio detection \textbf{[43]}, and video recognition \textbf{[44]}. It was designed by Sepp Hochreiter and Jürgen Schmidhuber in 1997 \textbf{[45]} to solve the vanishing gradient problem during optimization \textbf{[46]} in RNN by adding forget gate, input gate, and output gate into the hidden layer. 

Figure 1 shows the inner structure of the hidden state. There are three gates in each cell: forget gate, input gate, and output gate. Each gate consists of a $sigmoid$ activation plus a point-wise multiplication operation \textbf{[46]}. The forget gate evaluates how much information should be forgotten from the cell state, which constructs the long-term memory. The two inputs of the forget gate come from the output of the previous hidden state $H_{t-1}$, and the input of the current state $X_{t}$. We combine the two inputs with their weights plus their biases and then pass the result through the $sigmoid$ activation function (1) to estimate whether to keep or remove the information. Its result ranges from 0 to 1, where 0 indicates to forget the information completely, while 1 indicates to maintain that ultimately.

The input gate is to decide how much new information should be added to the cell. The $sigmoid$ function (2) does the same thing as it does in the forget gate. Then we create a candidate memory cell $\tilde{C}_{t} \in \mathbb{R}^{n\times h}$ and using $tanh$ function (3) to address how much memory in the old cell $\tilde{C}_{t-1} \in \mathbb{R}^{n\times h}$ should be maintained. We utilize $tanh$ activation function here because $tanh$ allows negative results, which means the cell state can either add or drop information. The red part of Figure 1 presents the updating of the cell state. By combining the multiplications between $F_t$ and $C_{t-1}$,  $I_t$ and $\tilde{C}_{t}$, we get the new cell state $C_t$ (4).

The output gate is to update the hidden state $H_t \in \mathbb{R}^{n\times h}$. Still, we employ a $sigmoid$ function (5) to figure out what information needs retain from the previous hidden state and the new input of the current state. Passing the updated cell state $C_t$ to $tanh$ function, we determine what to add or drop. Multiplying the $tanh$ function and the $sigmoid$ function, we finally get the new hidden state (6).

\begin{equation}
F_{t} = \sigma  (W_{xf}X_{t} + W_{hf} H_{t-1} + b_{f})
\end{equation}
\begin{equation}
I_{t} = \sigma  (W_{xf}X_{i} + W_{hi} H_{t-1} + b_{i})
\end{equation}
\begin{equation}
\tilde{C}_{t} = tanh (W_{xi}X_{t} + W_{hi} H_{t-1} + b_{i})
\end{equation}
\begin{equation}
C_{t} = F_{t} \odot C_{t-1} + I_t \odot \tilde{C}_{t}
\end{equation}
\begin{equation}
O_{t} = \sigma  (W_{xo}X_{t} + W_{ho} H_{t-1} + b_{o})
\end{equation}
\begin{equation}
H_{t} = O_{t} \odot tanh(C_{t})
\end{equation}

\begin{figure}[htp]
  \centering
  \includegraphics[width=0.9\linewidth]{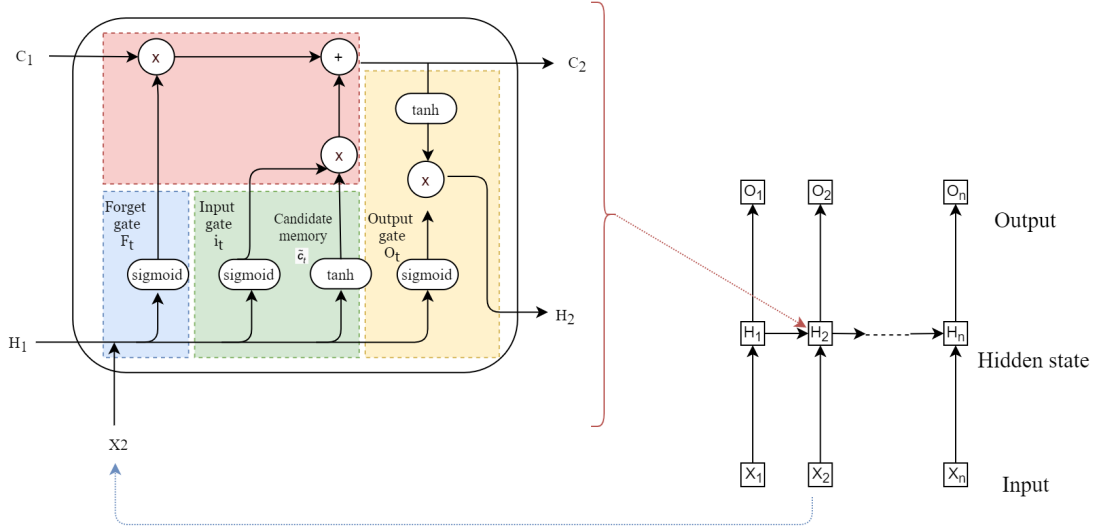}
  \caption{LSTM Structure}
  \label{fig:fig1.1}
\end{figure}

\subsection{Gated Recurrent Unit}
GRU was proposed by Cho et al. \textbf{[47]} in 2014. It also addresses the vanishing gradient problem of RNN. Unlike LSTM, however, it only has two gates: reset gate and update gate. The reset gate handles the short-term dependencies by deciding which past information should be dropped. Update gate is responsible for the long-term dependencies by controlling the amount of information passed to the future.  

 Assuming there are $h$ hidden units, $n$ examples, and $d$ inputs. Given the time step $t$, the input of the current state is $X_{t} \in \mathbb{R} ^ n \times d$, and the input from the previous hidden state is $H_{t-1} \in \mathbb{R}^n \times d$ . $X_{t}$ and $H_{t-1}$ are multiplied by their own weights $W_{xz} \in \mathbb{R} ^ n \times d$, $W_{hz} \in \mathbb{R} ^ h \times h$ respectively and then their summation is passed by a $sigmoid$ activation function to get the reset gate $R_{t} \in \mathbb{R} ^ n \times d$ ranging from 0 to 1 (7) and the update gate $Z_{t} \in \mathbb{R} ^ n \times d$ ranging from 0 to 1 (8). To get the candidate hidden state referred as the red part of Figure 2, we pass the summation to a $tanh$ activation function, and replacing $H_{t-1}$ by $H_{t-1}\times R_{t}$ to reduce the influence of previous states \textbf{[41]}. The yellow part of Figure 2 shows the process of getting the new $H_{t}$. We  need to incorporate the update gate, determine the extent of the influence from the previous state $H_{t-1}$, and the new candidate state $\tilde{H}_{t}$. Equation (10) shows the final updating step.

\begin{equation}
R_{t} = \sigma  (W_{xr}X_{t} + W_{hr} H_{t-1} + b_{z})
\end{equation}
\begin{equation}
Z_{t} = \sigma  (W_{xz}X_{t} + W_{hz} H_{t-1} + b_{z})
\end{equation}
\begin{equation}
\tilde{H}_{t} = tanh(W_{xh}X_{t} + W_{hh}(R_{t}H_{t-1}) + b_{z})
\end{equation}
\begin{equation}
H_{t} = Z_{t}H_{t-1} + (1-Z_{t})\tilde{H}_{t}
\end{equation}

\begin{figure}[htp]
  \centering
  \includegraphics[width=0.9\linewidth]{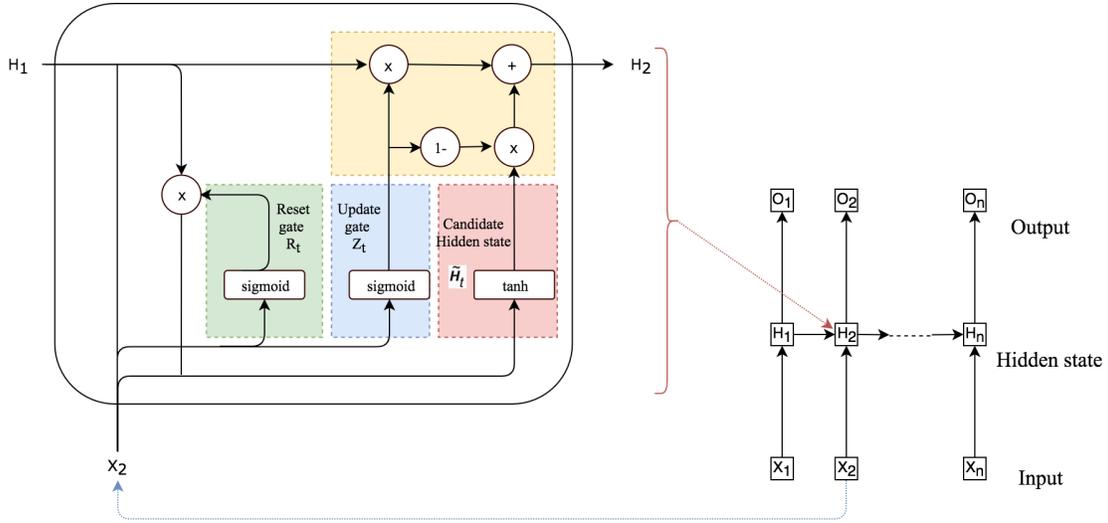}
  \caption{GRU Structure}
  \label{fig:fig2}
\end{figure}

\subsection{Bidirectional Recurrent Neural Network}
BiRNN was proposed to predict the target through both past and future information. As Figure 3 shows, it inserts an additional hidden layer in the RNN architecture for passing information in a backward direction, and then the output would be created by merging the information from two hidden layers, which is a combination of inputs from the opposite direction. To avoid vanishing gradient problem in RNN, researchers usually use bidirectional LSTM (BiSTM) and bidirectional GRU (BiGRU) instead of BiRNN. BiLSTM and BiGRU were initially designed for text, handwritten, or speech recognition, classification, and prediction. While recently, some researchers find it has excellent performance on 1-dimensional time series data forecasting, such as stock market prediction \textbf{[37]}, traffic speed prediction \textbf{[38]}, and salary prediction \textbf{[39]}
\begin{figure}[htp]
  \centering
  \includegraphics[width=0.5\linewidth]{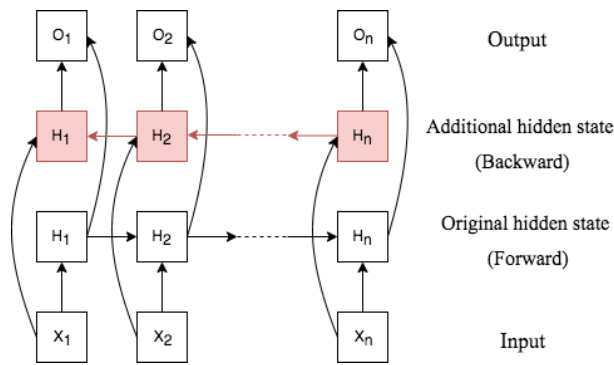}
  \caption{Bidirectional RNN Structure}
  \label{fig:fig3}
\end{figure}

\subsection{Convolutional Neural Network}
The CNN has achieved remarkable performance on images classification or recognition. Its success in two dimensions let some researchers explore 1D CNN on data regression problems. In the field of time series, its prediction only depends on local regions of independent variables rather than connecting to the input span over the long term. 1D CNN works well on extracting features from short fixed-length input of the whole data set and their locations have low relevance \textbf{[48]}. After the 1D CNN layer, we pass the result to a max pooling layer to prevent overfitting. It was developed to TCN with larger receptive field.

\subsection{Temporal Convolutional Network}
The ‘TCN’ in this paper specifically refers to the architecture proposed by the paper  \textit{An Empirical Evaluation of Generic Convolutional and Recurrent Networks for Sequence Modeling} published in 2014 \textbf{[40]}. After a series of experiments comparing with LSTM and GRU, the author claimed the TCN outperformed in a wide range of sequence modeling tasks.   

Its main characteristics include: 1) parallelism, 2) flexible receptive field size, 3) stable gradients, 4) low memory requirement for training,  and 5) flexible length of inputs. There are two different points between the TCN and CNN: 1) The output of TCN has the same length as the input; 2) There is no leakage from the future information to the past in TCN. To achieve the first point,  TCN uses 1D fully convolutional network architecture \textbf{[49]}. To get the second point, it applies causal convolutions. 

TCN employs dilated convolutions as the causal convolutions \textbf{[50]} to build a long effective history size, in other words, to get a large receptive field. Figure 4 provides an example of a dilated causal convolution with a kernel size of 2 and a dilation factor of [1,2,4]. TCN applied residual blocks \textbf{[51]} to keep the stabilization of the deep network. In each residual bock, there are two layers of dilated causal convolution, weight normalization \textbf{[52]}, rectified linear unit (ReLU) \textbf{[53]}, and spatial dropout \textbf{[54]}. Plus, a 1*1 convolution is added in each residual block to ensure the output size is the same as the input size.

\begin{figure}[htp]
  \centering
  \includegraphics[width=0.8\linewidth]{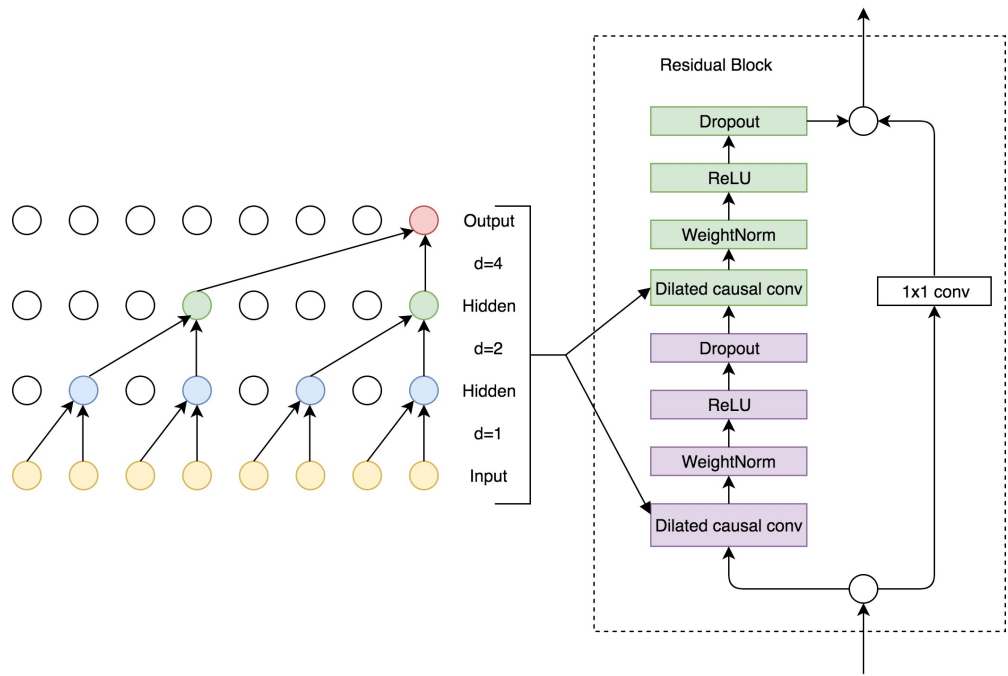}
  \caption{\footnotesize TCN Structure.  The left part is the dilated causal convolution with Kernel size=2 and dilation factors d=1,2,4. The right part is the residual block. The 1*1 convolution is to make sure the residual input and output have the same dimensions }
\end{figure}

\section{Experiment}

\subsection{Data and data preprocessing}
The data used in this experiment is obtained from a real-time water-quality monitoring station in the Yangtze River Basin of Jiangsu Province, Wuxi. The dissolved oxygen was recorded automatically each 30 min from 12.a.m. Aug 1st, 2012 to 9 p.m. May 8th, 2016, with a total of 64411 items. The distribution of the raw data is shown in Figure 5. \\
\begin{figure}[htp]
  \centering
  \includegraphics[width=0.8\linewidth]{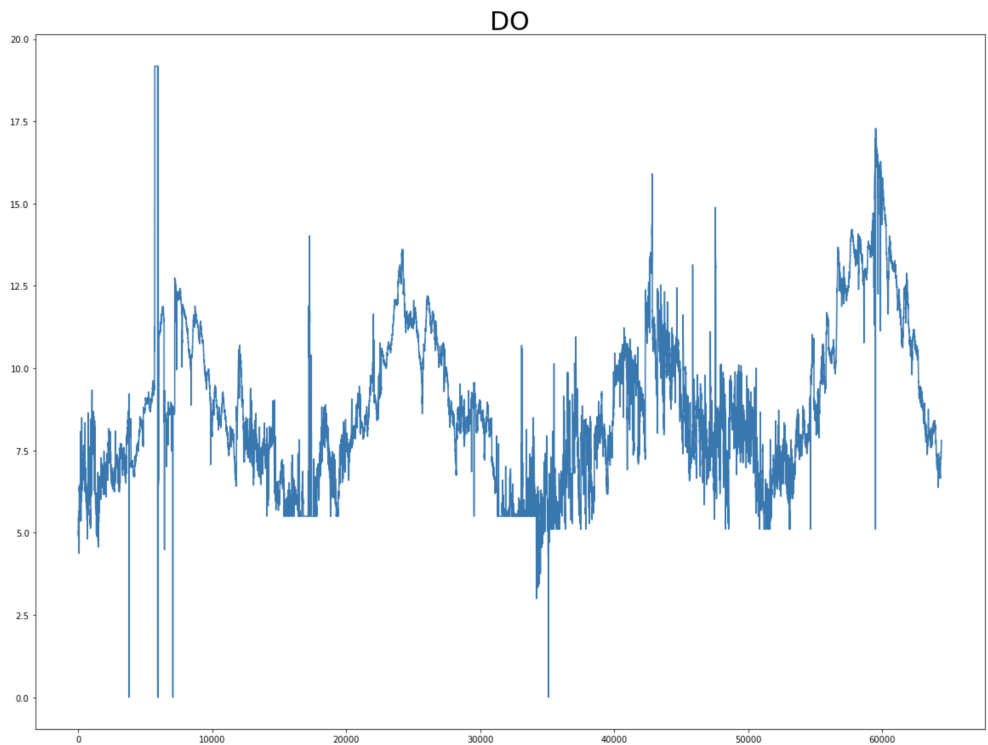}
  \caption{Raw data of dissolved oxygen}
  \label{fig:fig5}
\end{figure}
 Zeros are believed to be the errors made by monitor. There are 49 zeros among the whole items, which indicates that the rate of zero is 0.76\%. Considering its low percentage of occurrence and randomly dispersed distribution, we remove them to avoid their influence on the model training.  Tremendous amount of inputs with unscaled data will slow down the speed of learning and convergence of the deep learning model. Hence, we utilize the Min-Max strategy as the equation (11) to normalize the data and map its range from 0 to 1. Then We split the data into 90\% training set and 10\% testing set. Moreover, among the training data, we set 10\% validation set.
\begin{equation}
y = (x-min)/(max-min)
\end{equation}

\subsection{Model design}
\subsubsection{Loss function and Optimizer}
When we train the models in this study, we choose Mean Squared Error (MSE) as the loss function and Adam as the optimizer. Adam is created specifically for training deep learning models through combining the benefits of Adaptive Gradient algorithm (AdaGrad) \textbf{[55]} working well with sparse gradients, and Root Mean Square Propagation (RMSProp) \textbf{[56]} doing well on non-stationary objectives.
\subsubsection{Hypperparameters}
 We tune the hyperparameters of each model based on their keras packages. The amount of epochs of each model is 20. For LSTM, GRU, BiLSTM, and BiGRU, we set their units, batch size, validation split as 50, 128, and 0.1, respectively. For TCN, we set its dilations [1,2,4,8,16,32], with the kernel size of 3 and amount of filters as 4. For CNN, filters, kernel sizes, and validation split are 16, 3, 0.1.

\subsection{Model estimation} 
\subsubsection{Walk-forward Validation}
Many previous studies said little about their test model. That would easily make readers misunderstand the statistical metrics and conclusions. In this paper, We employ walk-forward validation to predict the observations of the next ten time steps, given the observations of the previous ten time steps. That is, using the previous five hours DO level to predict the DO level of the next five hours. There are two kinds of windows in walk-forward validation. In this paper, we choose the expanding window. As the first row of Figure 6 shows, we first input five hours of known data to predict data from the hour 6 to hour 10. The second row presents that we expand the input from five hours to ten hours, and then forecast the data for the next five hours. All predictions based on the model referred by the yellow squares in figure 6, is stored and finally be compared with the known value.
 
\begin{figure}[htp]
  \centering
  \includegraphics[width=0.8\linewidth]{test.pdf}
  \caption{Test Structure. }
  \label{fig:fig}
\end{figure}
\subsubsection{Statistical Metrics}
To test the predicting capability of the models, we select three statistical metrics and compare their values in the testing set. They are Root Mean Square Error (RMSE), Mean Absolute Error (MAE), and Coefficient of Determination {$R^2$}. For RMSE and MAE, closer to 0 means better performance, and for {$R^2$}, closer to 1 is better.  
\begin{equation}
 RMSE = \displaystyle\sqrt{\frac{1}{n}\sum_{t=1}^{n}(y_t - \hat{y_t})^2}
\end{equation}
\begin{equation}
 MAE = \displaystyle\frac{1}{n}\sum_{t=1}^{n}|y_t - \hat{y_t}| 
\end{equation}
\begin{equation}
\centering
R^2  = 1- \frac{\sum^{n-1}_{t=0}(y_t - \hat{y_t})^2}{\sum^{n-1}_{t=0}(y_t - \bar{y_t})^2}
\end{equation}
\subsubsection{Results}
Figure 7, 9 show that the prediction capability of all models falls dramatically after the time step of 2.5 hours and 4 hours. That is, the difference of reliability and accuracy between hour 3 and hour 2.5 is much larger than that between hour 2.5 and 2.0. It indicates that the attenuation of predicting performance is not a linear process.
According to Figure 7,8,9, GRU has the lowest average RMSE and MAE of each time step in the test set. GRU also has the highest average {$R^2$} in each time step. That presents strong evidence to support GRU outperforms other models in this study. We also find LSTM has the highest RMSE, MAE, while lowest {$R^2$} on average, which indicates its performance is the worst among the six models. 

\begin{figure}[htp]
     \centering
     \includegraphics[width=0.8\linewidth]{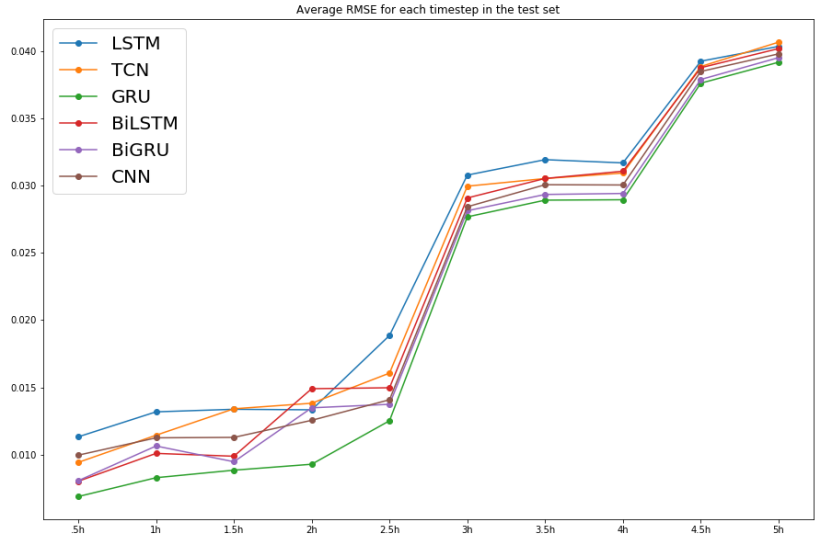}
     \caption{Average RMSE for each time step in the test set}
\end{figure}
\begin{figure}[htp]
     \centering
     \includegraphics[width=0.8\linewidth]{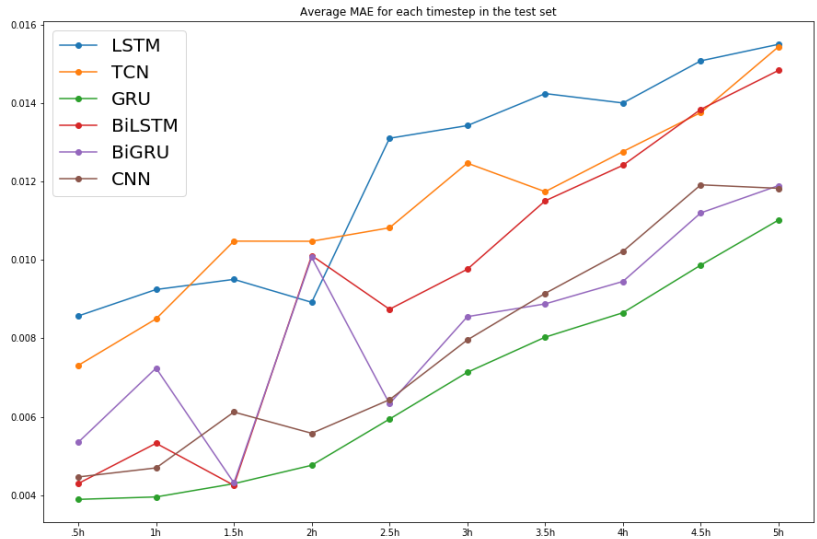}
     \caption{Average MAE for each time step in the test set}
\end{figure}
\begin{figure}[htp]
     \centering
     \includegraphics[width=0.8\linewidth]{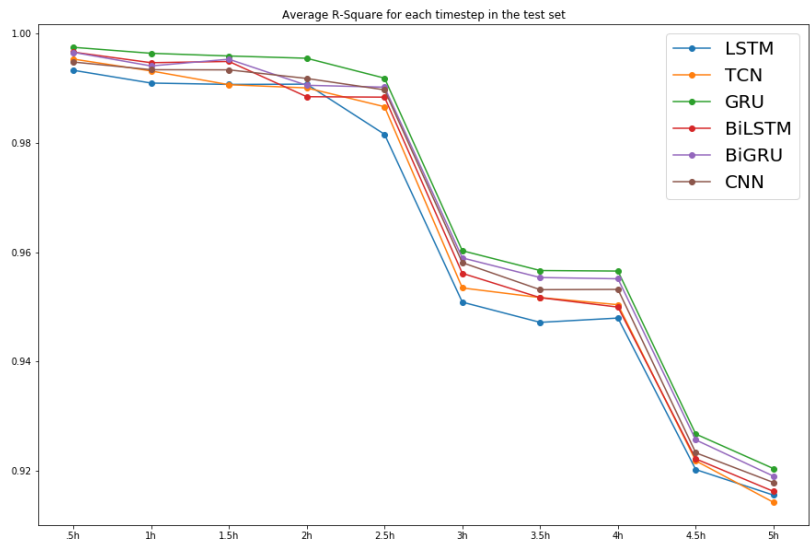}
     \caption{Average {$R^2$} for each time step in the test set}
\end{figure}

\section{Limitations and Future Works}
One limitation of this study is the quality of the raw data. Although we have dropped those zeros, it is hard for us to figure out those errors larger than zero. We see in Figure 5, there are several extreme values, and the values between 18000 to 20000 and 31000 to 32000 generate two horizontal bottom lines, which looks confusing. Although deep learning models have the advantage of dealing with those extreme values or error values, if the total error rate of the data is high, it is risk to believe the trained models are reliable enough. 

Another limitation is, although the expanding window test method is believed a scientific and excellent testing approach in time series forecasting. In the actual application, however,  during the model testing process, we need to consider to update the training model as the window moves forward because more known data is available. While that requires a much stronger computational power.

Finally, we tried some different combinations of hyperparameters and found their results did not change a lot. Hence, we choose big batch sizes, small units, and filters to accelerate the speed of training. Currently, there is still no gold standard of tuning the hyperparameter. While we believe a more rigorous proof or evidence might need to provide. Also, the result might be different if the length of input or output changes. The relationship among them needs more discussion in the future.

\section{Conclusion}
In this paper, we propose a sound comparison of predicting dissolved oxygen between multi-step deep learning models that predict DO levels of the next five hours, given the data of the last five hours. We analysed the performance of each model through walk-forward validation. It was observed that 1) GRU has the best performance, while LSTM has the lowest performance; 2) The attenuation of predicting performance of each model is not linear. The prediction capability of all models falls dramatically after 2.5 hours and 4 hours. That reminds the researchers to pay attention to the time nodes to prevent possible risks caused by the predictions with suddenly increased error rate.

\end{document}